# Measuring the 2D Vector Aspect of Momentum Using Only One Dimension

*Andrew Ferstl and Nathan Moore*
*Winona State University, Winona, MN*

Without the use of cameras to record 2D motion and an appropriate analysis tool, creating a laboratory activity for students to experience the vector nature of momentum can be challenging. Even with appropriate measurement tools, it is difficult to *predict* the final velocities of the objects in the system unless you know the impact parameter[i]. For accurately predicting the final velocities, the typical momentum experiment involves colliding gliders on an air track or colliding low-friction carts on a track. These one-dimensional experiments require the vector nature of momentum for correct analysis but do not require breaking the momentum vector into components.

In this work we present another experiment, using this aforementioned equipment, which only requires a stopwatch, ruler, and mass balance yet requires the two-dimensional analysis of the momentum vector.

## Lab Setup

The problem setup is illustrated in figure 1 and is similar to a typical homework problem found in physics books[ii]. A cart sits on a horizontal track at the bottom of an incline waiting for the sliding object to fall into it. The objective is to predict the final velocity of the cart/object system. The track we used was the Pasco 1.2 m Classic Dynamics System (ME-9435A). These come with accessories that allow the tracks to be attached to vertical posts to create the incline as well. The cart we used was the Pasco Collision Cart (ME-9454), which has a mass of 500 g (250 g carts are also available). In order to measure the cart velocity with a stopwatch at $\theta \approx 30^o$ and $d \approx 30 cm$ we used a 200g mass which can be found in Pasco's Mass and Hanger Set (ME-8979). Other types of track, cart, and mass sets are available from different manufacturers. The toughest part of the set up is to get the mass to slide into the cart without bouncing out (so that the collision is perfectly inelastic) or bouncing the cart of the track. You may also wish to pad the cart to make a better "trap" for the sliding mass. A successful lab also requires knowing the ramp/box coefficient of sliding friction, $\mu_k$, which could be established through a previous lab.

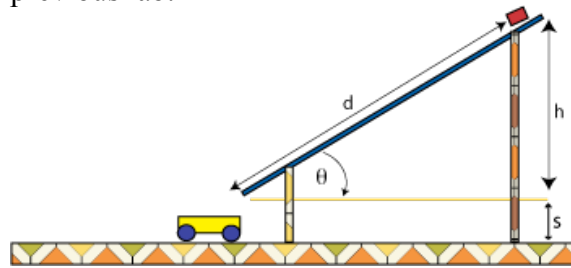

**Figure 1. This figure is generally presented to students before they come to lab.**

## The Prediction

Before coming to lab, we ask our students to work towards a conceptual and analytic understanding of the problem. In our experience[iii], this pre-lab work, or "prediction" significantly improves the lab experience for students in that they seem to learn comparatively more physics and do not seem to be so quickly frustrated when they encounter problems in the lab. Our students are encouraged to work on the prediction in cooperative groups before coming to lab, and continue with those groups in the lab period, which also reduces frustration and increases student learning.

This experiment involves a collision, so our analysis will use the momentum principle to describe the system before and after the collision. This requires calculating the velocity of the sliding mass, which we will now call the "box" (for reasons that will become apparent later), before it hits the cart. Since we are given the distance that the box slides down the ramp and we're looking for speed (the direction is implied by the ramp) at the end of the ramp, $v_{end}$, it is natural to use the energy principle. Take the system to be the box and the earth so that we can express the work done by gravity as a potential energy. In this case, friction will convert mechanical energy into thermal energy. If the box's mass is $m_b$ and the it slides a distance, $d,$ along the track, the energy principle[iv], $E_{transfer} = E_f - E_i$, using the point-particle model, becomes

$$-f_k d = \frac{1}{2} m_b v_{end}^2 - m_b g h \quad (1)$$

We still need to quantify the frictional force. We can use the relation between the frictional force and the force from the ramp (usually called the Normal force):

$$f_k = \mu_k F_{ramp} \quad (2)$$

where we can find $F_{ramp}$ by applying the momentum principle, $\vec{F}_{ext} \Delta t = \Delta \vec{p}$, in the direction perpendicular to the ramp and recognizing that the change in momentum in that direction is zero.

$$F_{ramp} - m_b g \cos\theta = 0 \quad (3)$$

If Eqs (1), (2), and (3) are combined along with some geometry to relate d to h, we can solve for the speed of the box as it leaves the ramp.

$$v_{end} = \sqrt{2gd(\sin\theta - \mu_k \cos\theta)} \quad (4)$$

Once the box leaves the ramp there is no force in the horizontal direction. Therefore, the momentum principle dictates that the horizontal velocity remains constant from the point it leaves the ramp until it hits the cart.

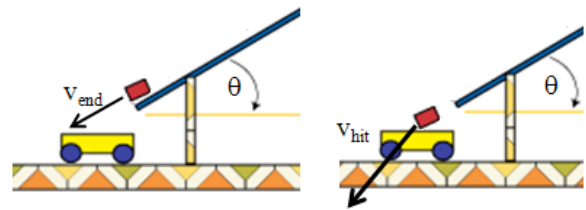

$$v_{hit,x} = v_{end,x} = v_{end} \cos\theta \quad (5)$$

The box collides with the cart inelastically, so proper analysis necessitates momentum. The analysis requires two-dimensions, because the box travels with vertical and horizontal momentum before the collision but there is only a horizontal momentum after the collision. Now, taking the system to be the box and the cart, then in the time interval of immediately before the collision to immediately after, there are no *horizontal* impulses on the system. Accordingly, the momentum principle becomes (where we have taken the + x axis to point horizontally in the direction of motion):

$$0 = (m_{cart} + m_b) v_f - m_b v_{hit,x} \quad (6)$$

or,

$$v_f = \frac{m_b}{m_b + m_{cart}} v_{hit,x} \quad (7)$$

In the vertical direction, the floor does exert a force so the momentum does change. The momentum principle says:

$$F_{floor}\Delta t = m_b v_{hit,y} \quad (8)$$

where $F_{floor}$ is the force the floor exerts on the system for a time interval, $\Delta t$. Although the vertical momentum is not strictly required for an accurate prediction, this corner of the problem can be useful when talking about the safety of the box (or delicate package) as it moves from the ramp to the cart. It is also useful when discussing why only the horizontal component of momentum is useful for predicting the final velocity.

With these physical ideas, one can use Eqs (4), (5), and (7) to predict the cart/box system's final velocity:

$$v_f = \frac{m_b \cos\theta}{m_b + m_{cart}} \sqrt{2gd(\sin\theta - \mu_k \cos\theta)} \quad (9)$$

**Suggestions for Implementation**

In this lab, one can easily vary the release position of the box on the ramp, $d$. The final speed of the cart, $v_f$, is proportional to $\sqrt{d}$, so for the purposes of plotting and fitting the data, a plot of $(v_f)^2$ versus $d$ might be the best way to see that the prediction equation is valid over a range of inputs. That said, when students are given the opportunity to create their own verification of the prediction, the results can be quite interesting and powerful for student learning.

Student generated data is compared to this prediction in figure 2:

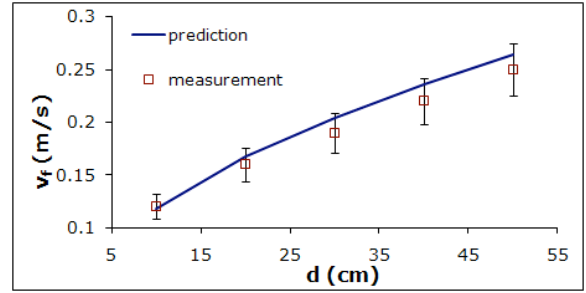

**Figure 2, Student data is compared to the prediction developed in this article at an angle of 27°.**

**Lab Pedagogy**

We present the lab problem to the students a few days before the lab is scheduled. Before coming to the lab period, we expect that students will have read through the lab problem, thought about how to apply the physics they know, and hopefully will have made an analytical prediction that will allow them to be well-prepared for lab. At the beginning of the lab period, the instructors solicit student ideas on what the solution is and the best way to approach the problem. Without correcting any student alternative conceptions (see below), they are allowed to collect data to compare their results with the prediction. The accuracy of the lab is very important if students are to discover, on their own, if their prediction is correct or not. The discussion of the problem's solution is continued at the end of the lab period so that the students have a chance to resolve possible misconceptions about the pertinent physics.

To set the context and relevance of the lab a possible story could be (which is why we called the sliding mass a "box"):

*Your summer job at a local delivery service involves moving packages from the sorting room to the truck loading room. Your boss continually reminds the crew that any successful efficiency improvements will be financially rewarded. The mechanics of your job include putting a labeled box into the corresponding chute. The box slides down a smooth incline, and then drops into a conveyor cart that brings the package to the appropriate delivery truck. To prevent the packages from being damaged, the boxes land on a foam pad. After two weeks of work, you realize that perhaps the box's motion could be used to propel the cart to the loading ramp. The key in your solution will be to understand the relationship between the final speed of the box/cart and the ramp the box slides down.*

*Lab Question: For the setup shown in figure 1, predict the final velocity of the cart after the box lands in it. Identify the physical parameters that are important. Make some realistic assumptions about the system, i.e., the package slides down the ramp without rolling, it slides down with a constant frictional drag force, etc.*

**Common Student Misconceptions**

One common student misconception is to ignore the conversion of kinetic energy into thermal energy and sound which is implicit in the box's inelastic collision with the cart. The students' prediction equation then might look like the following, taking the initial state to be the box at the top and the final state is the moving cart:

$$-f_k d = \frac{1}{2}(m_b + m_{cart})v_f^2 - m_b gh \quad (10)$$

Using Eqs (2) and (3) and solving for the speed gives:

$$v_f = \sqrt{\frac{m_p}{m_p + m_{cart}} 2gd(\sin\theta - \mu_k \cos\theta)} \quad (11)$$

Comparing with the correct prediction, Eq (9), one sees that this should visibly contradict experimental data. In addition to being too large due to the absence of $\cos\theta$, this prediction also is missing a factor of $\sqrt{\frac{m_p}{m_p + m_{cart}}}$, which leads to an over-estimation of the cart's final velocity. This misconception can be resolved conceptually: If the student assumes that the collision is elastic (i.e. kinetic energy is conserved). then the energy normally converted to heat, sound, and deformations will lead to a higher than measured final cart speed.   This misconception is usually accompanied by the students using the total change in height, h+s, as opposed to just the change in height along the ramp.  As discussed previously, the distance from the end of the ramp to the cart, s, is irrelevant for the purpose of predicting the final cart/box velocity.

A more sophisticated misconception is ignoring the 2-d aspect of the collision. In this misconception, a student might successfully derive Eq (4), and understand that they need use momentum conservation for the collision. The misconception however begins when they forget that momentum, unlike energy, has direction. The resulting final prediction differs only from the correct prediction, Eq (9), by a factor of $\cos\theta$.

$$v_f = \frac{m_p}{m_p + m_{cart}}\sqrt{2gd(\sin\theta - \mu_k \cos\theta)} \quad (12)$$

As $\cos\theta \approx 1$ for small angles, it is important to have a significant incline to the track to make this second misconception visible when compared to data.

**Conclusions**

We have successfully implemented a lab that requires the vector nature of momentum without requiring equipment for analyzing 2-D motion. The lab allows students to explore their conceptions of momentum and energy. Since energy and momentum are often used together to describe collisions, it is easy for students to forget that momentum is a vector (because energy is a scalar). Besides the aforementioned misconceptions, some students are also concerned with the fact that the box will bounce when it hits the cart which they interpret to mean that momentum for the system is not conserved (in the horizontal direction).

We do not have any quantitative data (for example, pre/post testing) to suggest that this lab enhances student learning but the lab does permit the discussion of several student misconceptions about energy and momentum which we feel is an important step toward understanding.

---

[i] Irvin A. Miller, "Two-dimensional collisions using pendulums," *Phys.Teach.* **27**, 207-209 (1989). This 2D experiment allows students to see the vector nature of momentum but unless the impact parameter of the colliding marbles is known, it is impossible to predict the final velocities (and hence final positions) of the marbles.

[ii] See for example, Young and Freedman, *University Physics,* 11th ed, (Pearson/Addison Wesley, 2004) problem 81, pg 323.

[iii] The lab problem structure we use is heavily influenced by our time at the University of Minnesota, whose physics department uses Problem Solving Labs, which are described on the web at, http://groups.physics.umn.edu/physed/Research/PSL/pslintro.html .

[iv] Our notation for energy conservation is similar to the language used in Chabay and Sherwood's, *Matter and Interactions I*, 2nd ed (Wiley, 2007) where $E_{transer}$ represents both work and heat.